\documentstyle[12pt]{article}
\begin{document}
\begin{center}
{\Large {\bf Bi-Hamiltonian structure as a shadow of non-Noether symmetry}}\\
\vspace*{5mm}
{\bf George Chavchanidze}
\end{center}
\begin{abstract}
In the present paper correspondence between non-Noether symmetries and bi-Hamiltonian structures
is disscussed. We show that in regular Hamiltonian systems presence of the global bi-Hamiltonian
structure is caused by symmetry of the space of solution. As an example well known bi-Hamiltonian
realisation of Korteweg- De Vries equation is disscussed. 
\end{abstract}

Noether theorem, Lutzky's theorem, bi-Hamiltonian formalism and bidifferential calculi are often used 
in generating conservation laws and all
this approaches are unified by the single idea - to construct conserved quantities out of some invariant
geometric object (generator of the symmetry - Hamiltonian vector field in Noether theorem, 
non-Hamiltonian one in Lutzky's approach, closed 2-form in bi-Hamiltonian formalism and auxiliary
differential in case of bidifferential calculi). There is close relationship between later three approaches.
Some aspects of this relationship has been uncovered in [3-4]. In the present paper it is
discussed how bi-Hamiltonian structure can be interpreted as a manifestation of symmetry of space of
solutions. Good candidate for this role is non-Noether symmetry. Such a symmetry is a group of
transformation that maps the space of solutions of equations of motion onto itself, but unlike the
Noether one, does not preserve action. \\
In the case of regular Hamiltonian system phase space is equipped with symplectic form $\omega $
(closed $d\omega = 0$ and nondegenerate $i_{X}\omega = 0 \rightarrow X = 0$ 2-form) and time
evolution is governed by Hamilton's equation
\begin{eqnarray}
i_{X_{h}}\omega + dh = 0 \label{eq:ham}
\end{eqnarray}
where $X_{h}$ is vector field tangent to solutions 
$X_{h} = \Sigma _{i} \dot p_{i}\partial _{p_{i}} + \dot q_{i}\partial _{q_{i}}$ and
$i_{X_{h}}\omega $ denotes contraction of
$X_{h}$ and $\omega $. Vector field is said to be (locally) Hamiltonian if it preserves $\omega $.
According to the Liouville's theorem $X_{h}$ defined by (\ref{eq:ham}) automatically preserves $\omega $ (indeed
$L_{X_{h}}\omega = di_{X_{h}}\omega + i_{X_{h}}d\omega = - ddh = 0$).\\
One can show that group of transformations of phase space generated by any non-Hamiltonian vector
field $E$
\begin{eqnarray}
g(a) = e^{aL_{E}}
\end{eqnarray}
does not preserve action
\begin{eqnarray}
g_{*}(A) = g_{*}(\int pdq - hdt) = \int g_{*}(pdq - hdt)\neq 0
\end{eqnarray}
because $d(L_{E}(pdq - hdt)) = L_{E}\omega - dE(h)\wedge dt \neq 0$ (first term in rhs does not vanish
since $E$ is non-Hamiltonian and as far as $E$ is time independent $L_{E}\omega $ and 
$dE(h) \wedge dt$ are linearly independent 2-forms). As a result every non-Hamiltonian vector field $E$
commuting with $X_{h}$ leads to the non-Noether symmetry (since $E$ preserves vector field tangent
to solutions $L_{E}(X_{h}) = [E , X_{h}] = 0$ it maps the space of solutions onto itself). Any such
symmetry yields the following integrals of motion [1-2], [4-5]
\begin{eqnarray}
l_{k} = Tr(R^{k}) ~~ k = 1,2 ... n
\end{eqnarray}
where $R = \omega ^{-1}L_{E}\omega $ and $n$ is half-dimension of phase space.\\
It is interesting that for any non-Noether symmetry, triple $(h, \omega , \omega _{E})$ carries 
bi-Hamiltonian structure ( \S 4.12 in [6], [7-9]). Indeed $\omega _{E}$ is closed 
($d\omega _{E} = dL_{E}\omega = L_{E}d\omega = 0$) and invariant 
($L_{X_{h}}\omega _{E} = L_{X_{h}}L_{E}\omega = L_{E}L_{X_{h}}\omega = 0$) 
2-form (but generic $\omega _{E}$ is degenerate). So every non-Noether
symmetry quite naturally endows dynamical system with bi-Hamiltonian structure. \\
Now let's discuss how non-Noether symmetry can be recovered from bi-Hamiltonian system. Generic 
bi-Hamiltonian structure on phase space consists of Hamiltonian system $h, \omega $ and auxiliary
closed 2- form $\omega ^{\bullet }$ satisfying $L_{X_{h}}\omega ^{\bullet } = 0$. Let us call it global 
bi-Hamiltonian structure whenever $\omega ^{\bullet }$ is exact (there exists 1-form $\theta ^{\bullet }$ such that
$\omega ^{\bullet } = d\theta ^{\bullet }$) and $X_{h}$ is (globally) Hamiltonian vector field with respect to
$\omega ^{\bullet }$ ($i_{X_{h}}\omega ^{\bullet } + dh^{\bullet } = 0$). In the local coordinates 
$\theta ^{\bullet } = \theta _{i}^{\bullet }dz^{i}.$
As far as $\omega $ is nondegenerate there exists vector field 
$E^{\bullet } = E^{\bullet i}\partial _{z^{i}}$ such that 
$i_{E^{\bullet }}\omega = \theta ^{\bullet }$ (in local coordinates 
$E^{\bullet i} = (\omega ^{-1})^{ij}\theta _{j}^{\bullet }$). 
By construction
\begin{eqnarray}
L_{E^{\bullet }} \omega = \omega ^{\bullet }
\end{eqnarray}
Indeed $L_{E^{\bullet }}\omega = di_{E^{\bullet }}\omega + i_{E^{\bullet }}d\omega 
= d\theta ^{\bullet } = \omega ^{\bullet }$.\\
And
\begin{eqnarray}
i_{[E^{\bullet },X_{h}]}\omega = 
L_{E^{\bullet }}(i_{X_{h}}\omega ) - i_{X_{h}}L_{E^{\bullet }}\omega 
= - d(E^{\bullet }(h) - h^{\bullet }) = - dh'
\end{eqnarray}
In other words $[X_{h} , E^{\bullet }]$ is Hamiltonian vector field, i. e., $[X_{h} , E] = X_{h'}$. So
$E^{\bullet }$ is not generator of symmetry since it does not commute with $X_{h}$ but one can
construct (locally) Hamiltonian counterpart of $E^{\bullet }$ (note that $E^{\bullet }$ itself is 
non-Hamiltonian) - $X_{g}$ with 
\begin{eqnarray}
g(z) = \int _{0}^{t}h'dt . \label{eq:gz}
\end{eqnarray}
Here integration along solution of Hamilton's equation, with fixed origin and end point in $z(t) = z$,
is assumed. Note that (\ref{eq:gz}) defines $g(z)$ only locally and, as a result, $X_{g}$ is a locally
Hamiltonian vector field, satisfying, by construction, the same commutation relations as 
$E^{\bullet }$ (namely $[X_{h} , X_{g}] = X_{h'}$). 
Finally one recovers generator of non-Noether symmetry - non-Hamiltonian vector field 
$E = E^{\bullet } - X_{g}$ commuting with $X_{h}$ and satisfying
\begin{eqnarray}
L_{E}\omega = L_{E^{\bullet }}\omega - L_{X_{g}}\omega = L_{E^{\bullet }}\omega = \omega ^{\bullet }
\end{eqnarray}
(thanks to Liouville's theorem $L_{X_{g}}\omega = 0$). So in case of regular Hamiltonian system every
global bi-Hamiltonian structure is naturally associated with (non-Noether) symmetry of space of
solutions. \\

{\bf Example 1} As a toy example one can consider free particle
\begin{eqnarray}
h = \frac{1}{2} \sum_{i} p_{i}^{2}~~~~\omega = \sum_{i} dp_{i}\wedge dq_{i}
\end{eqnarray}
this Hamiltonian system can be extended to the bi-Hamiltonian one
\begin{eqnarray}
h, \omega , \omega ^{\bullet } = \sum_{i} p_{i}dp_{i}\wedge dq_{i}
\end{eqnarray}
clearly $d\omega ^{\bullet } = 0$ and $X_{h} = \sum_{i} p_{i}\partial _{q_{i}}$ preserves 
$\omega^{\bullet }$. Conserved quantities $p_{i}$ are associated with this simple 
bi-Hamiltonian structure.
This system can be obtained from the following (non-Noether) symmetry (infinitesimal form)
\begin{eqnarray}
q_{i} ~~\rightarrow ~~(1 + ap_{i})q_{i}
\\
p_{i} ~~\rightarrow ~~(1 + ap_{i})p_{i}
\end{eqnarray}
generated by $E = \sum_{i} p_{i}q_{i}\partial _{q_{i}} + \sum_{i} p_{i}^{2} \partial _{p_{i}}$
\\

{\bf Example 2.} The earliest and probably the most well known bi-Hamiltonian structure is the one
discovered by F. Magri and assosiated with Korteweg- De Vries integrable hierarchy. The KdV equation
\begin{eqnarray}
u_{t} + u_{xxx} + uu_{x} = 0
\end{eqnarray}
(zero boundary conditions for $u$ and its derivatives are assumed) appears to be Hamilton's equation
\begin{eqnarray}
i_{X_{h}}\omega + dh = 0
\end{eqnarray}
where $X_{h} =  \int _{- \infty }^{+ \infty } dx u_{t} \partial _{u}$ (here $\partial _{u}$ 
denotes variational derivative with respect to the field $u(x)$) is the vector field tangent to the
solutions,
\begin{eqnarray}
\omega =  \int _{- \infty }^{+ \infty } dx ~du\wedge dv
\end{eqnarray}
is the symplectic form (here $v = \partial _{x}^{-1}u$) and the function
\begin{eqnarray}
h =  \int _{- \infty }^{+ \infty } dx (\frac{u^{3}}{3} - u_{x}^{2})
\end{eqnarray}
plays the role of Hamiltonian. This dynamical system possesses non-trivial symmetry - one-parameter
group of non-cannonical transformations $g(a) = e^{L_{E}}$ generated by the non-Hamiltonian vector
field
\begin{eqnarray}
E =  \int _{- \infty }^{+ \infty } dx (u_{xx} + \frac{1}{2}u^{2})\partial _{u} + X_{F}
\end{eqnarray}
here first term represents non-Hamiltonian part of the generator of the symmetry, while the second one
is its Hamiltonian counterpart assosiated with
\begin{eqnarray}
F =  \int _{- \infty }^{+ \infty }(\frac{1}{12}u^{2}v + \frac{1}{4}\partial _{x}^{- 1}(\frac{u^{3}}{3} - u_{x}^{2}) + \frac{3}{4}v\frac{l_{3}}{l_{2}})dx
\end{eqnarray}
($l_{2,3}$ are defined in (\ref{eq:integrals}). The physical origin of this symmetry is unclear, however the
symmetry seems to be very important since it leads to the celebrated infinite sequence of conservation
laws in involution:
\begin{eqnarray}
l_{1} =  \int _{- \infty }^{+ \infty } u dx \nonumber \\
l_{2} =  \int _{- \infty }^{+ \infty } u^{2} dx \nonumber \\
l_{3} =  \int _{- \infty }^{+ \infty } (\frac{u^{3}}{3} - u_{x}^{2}) dx \nonumber \\
l_{4} =  \int _{- \infty }^{+ \infty } (\frac{5}{36}u^{4} - \frac{5}{3}uu_{x}^{2} + u_{xx}^{2}) dx
\label{eq:integrals} \\
... \nonumber
\end{eqnarray}
and ensures integrability of KdV equation. Second Hamiltonian realization of KdV equation discovered
by F. Magri [7]
\begin{eqnarray}
i_{X_{h^{\bullet }}}\omega ^{\bullet } + dh^{\bullet } = 0
\end{eqnarray}
(where $\omega ^{\bullet } = L_{E}\omega $ and $h^{\bullet } = L_{E}h$) is a result of 
invariance of KdV under aforementioned transformations $g(a)$.\\

{\bf Acknowledgements.} {\it Author is grateful to Z. Giunashvili for constructive discussions and to G.
Jorjadze for support. This work was supported by INTAS (00-00561) and Scholarship from World
Federation of Scientists.} \\
 
Author's Address:\\
A. Razmadze Mathematical Institute \\
1 Aleksidze St. , Tbilisi 380093, Georgia \\
E-mail: gch@rmi.acnet.ge
\end{document}